\documentclass[aps, prd, superscriptaddress, twocolumn, notitlepage, 10pt, nofootinbib]{revtex4}

\usepackage{amssymb, amsmath} 
\usepackage{stmaryrd}

\usepackage{yfonts}

\usepackage{bbm} 

\usepackage[english]{babel}
\usepackage{slashed}
\usepackage{cancel}

\usepackage{graphicx}
\usepackage{color}

\usepackage{datetime}
\settimeformat{ampmtime}

\usepackage{float}

\usepackage[bookmarks,breaklinks,plainpages=false,pdfpagelabels]{hyperref} 
	\hypersetup{linkcolor=red,citecolor=blue,filecolor=dullmagenta,urlcolor=darkblue} 

\definecolor{grey}{rgb}{0.4,0.4,0.4}
\definecolor{lightgrey}{rgb}{0.6,0.6,0.6}
\definecolor{dullmagenta}{rgb}{0.4,0,0.4}
\definecolor{darkblue}{rgb}{0,0,0.4}
\definecolor{orange}{rgb}{1,0.5,0}
\definecolor{lightbrown}{rgb}{0.75,0.5,0.25}
\definecolor{tan}{cmyk}{0.14,0.42,0.56,0}
\definecolor{djunglegreen}{cmyk}{0.99,0,0.52,0}
\definecolor{lightgreen}{rgb}{0,1,0}
\definecolor{olivegreen}{cmyk}{0.64,0,0.95,0.40}
\definecolor{midgreen}{rgb}{0.0,0.675,0.0}
\definecolor{darkgreen}{rgb}{0,0.5,0}

\newcommand{\qq}{\qquad}

\newcommand{\qqqq}{\qquad\qquad}
\newcommand{\vs}{\vspace}

\renewcommand{\.}{\hspace{0.5mm}}

\newcommand{\ra}{\ensuremath{\rightarrow}}

\newcommand{\Vrm}{\ensuremath{\mathrm{V}}}

\newcommand{\erm}{\ensuremath{\mathrm{e}}}

\newcommand{\irm}{\ensuremath{\mathrm{i}}}

\newcommand{\Hcal}{\ensuremath{\mathcal{H}}}

\newcommand{\Rcal}{\ensuremath{\mathcal{R}}}

\renewcommand{\d}{\ensuremath{\mathrm{d}}}

\newcommand{\defas}{\mathrel{\mathop :}=} 
\newcommand{\asdef}{=\mathrel{\mathop :}} 

\newcommand{\eg}{e.g.}
\newcommand{\ie}{i.e.}

\newcommand{\hc}{{\rm H.c.}}

\newcommand{\wrt}{w.r.t.}
\newcommand{\cf}{cf.}

\makeatletter
 \def\ifundefined#1{\expandafter\ifx\csname#1\endcsname\relax}
 \ifundefined{default@color}
  \let\default@color=\current@color
 \fi
\makeatother

\newcommand{\beq}{\begin{equation}}
\newcommand{\eeq}{\end{equation}}
\newcommand{\bea}{\begin{eqnarray}}
\newcommand{\beas}{\begin{eqnarray*}}

\newcommand{\eea}{\end{eqnarray}}
\newcommand{\eeas}{\end{eqnarray*}}

\newcommand{\fett}[1]{\boldsymbol{#1}}

\newcommand{\dd}{{\rm{d}}}

\newcommand{\be}{\begin{equation}}
\newcommand{\ee}{\end{equation}}

\newcommand{\nabq}{\fett{\nabla}_{\!\fett{q}}}
\newcommand{\nabx}{\fett{\nabla}_{\!\fett{x}}}

\begin{document}

\title{Astrophysical Bose-Einstein Condensates and Superradiance}

\date{\formatdate{\day}{\month}{\year}, \currenttime}

\author{Florian K{\"u}hnel}
	\email{florian.kuhnel@fysik.su.se}
	\affiliation{The Oskar Klein Centre for Cosmoparticle Physics,
			Department of Physics,
			Stockholm University,
			AlbaNova,
			106\.91 Stockholm,
			Sweden}

\author{Cornelius Rampf}
	\email{cornelius.rampf@port.ac.uk}
	\affiliation{Institute of Cosmology and Gravitation,
			University of Portsmouth,
			Portsmouth PO1 3FX,
			United Kingdom}
	\affiliation{Max-Planck-Institute for Gravitational Physics (Albert-Einstein-Institute),
			14476 Potsdam-Golm,
			Germany}

\begin{abstract}
We investigate gravitational analogue models to describe slowly rotating objects (\eg, dark-matter halos, or boson stars) in terms of Bose-Einstein condensates, trapped in their own gravitational potentials. We begin with a modified Gross-Pitaevskii equation, and show that the resulting background equations of motion are stable, as long as the rotational component is treated as a small perturbation. The dynamics of the fluctuations of the velocity potential are effectively governed by the Klein-Gordon equation of a ``Eulerian metric'', where we derive the latter by the use of a relativistic Lagrangian extrapolation. Superradiant scattering on such objects is studied. We derive conditions for its occurence and estimate its strength. Our investigations might give an observational handle to phenomenologically constrain Bose-Einstein condensates.
\end{abstract}

\maketitle

\section{Introduction}
\label{sec:Introduction}

The history of superradiance has possibly started with the discovery of the (inverse) Compton effect in 1923 by A.~H.~Compton \cite{Compton:1923zz}, and has since then stimulated various works in different fields of physics and mathematics. The specific phenomenon of inertial superradiance due to the superluminal motion of a (charged) object through some medium goes back to investigations by P.~A.~Cherenkov in 1934 \cite{Frank:1937fk, Ginzburg:1945zz}. Since these pioneering works which were both awarded with a Nobel Prize, also the idea of rotational superradiance arose, \ie, where a wave is inelastically scattered on a rotating rigid object such that its rotational momentum is transferred to the energy of the wave \cite{Dicke:1954zz} (for an overview of superradiant phenomena see, \eg, \cite{Bekenstein:1998nt}). In particular, investigations by Ya.~B.~Zel'dovich in 1972 then showed \cite{zeldovich19971, zeldovich1972},\footnote{See also the contributions to this subject coming from C.~W.~Misner \cite{Misner:1972kx} and W.~G.~Unruh \cite{Unruh:1974bw}.} that superradiance should be also spontaneously emitted by rotating black holes, where the space-time geometry can be described by a Kerr metric. Initially, the argument of superradiance was based on purely quantum-mechanical considerations, but it was Bekenstein in 1973 who realised that the argument should also hold in the classical sense to satisfy Hawking's classical horizon area theorem for black holes \cite{Bekenstein:1973mi, Bekenstein:1998nt}.

Stimulated by these works but also from the idea that black holes could radiate Hawking radiation \cite{Hawking:1974sw}, W.~G.~Unruh proposed in 1980 that the very occurence of Hawking radiation is not related with generic properties of gravity, but just the result of evolving quantised perturbations at an (apparent) event horizon \cite{Unruh:1980cg, Visser:2001kq}. As a consequence, Hawking radiation should also become apparent at a sonic horizon when a fluid flow becomes transonic. The idea of analogue models of gravity, often dubbed acoustic black holes, was born.\footnote{For a recent review on analogue gravity see Ref.~\cite{Barcelo:2005fc}.} Since then, various efforts have been done to study Bose-Einstein condensates (BEC) in the laboratory to establish and to study analogue event horizons (\cf~\cite{Barcelo:2005fc, Barcelo:2001ca} and references therein).

Soon it became also clear that not only Hawking radiation but also superradiance should arise at acoustic black holes, simply because superradiance is the consequence of forming some horizon and an ergosphere \cite{Basak:2002aw,Basak:2003uj} (provided that appropriate boundary conditions are established \cite{Richartz:2009mi}), and thus has in general nothing to do with a space-time singularity, neither with gravity in specific. Thus, analogous to Unruh's idea, the analogue scenario for extracting angular momentum from Kerr black holes would be a supersonic and a (mildly) rotating fluid flow.

In recent years, it has become very fruitful not only to study BECs in the laboratory but also to interpret macroscopic objects---such as (primordial) black holes \cite{Khlopov:1985jw,Dvali:2011aa, Dvali:2012en, Kuhnel:2013uxa}, neutron\./\.boson stars \cite{Wang:2001wq, Gruber:2014mna, Kleihaus:2011sx}, white dwarfs \cite{Benvenuto:2011fj}, and\./\.or dark matter halos as BECs \cite{Boehmer:2007um, Sikivie:2009qn, Chavanis:2011gm}.

Here we follow a similar objective, although our approach is to some extent new. We shall begin with a modified Hamiltonian of the BEC with gravitational self-interactions, and we include a rotational kinetic term which is greatly suppressed \wrt~its non-rotational kinetic term (see the following Sec.~\ref{sec:System-and-Equations-of-Motion}). In Sec.~\ref{sec:Stability} we then show that the resulting equations of motion are indeed stable, and we solve for the resulting density and velocity distribution of the BEC.

Generally, the inclusion of a rotational term in the Hamiltonian is known in studies related with generating vortex states in rapidly rotating BECs (\cf~Refs.~\cite{Yu:2002sz,Zinner:2011if, RindlerDaller:2011kx, Federici:2005ty}), such as in superfluid Helium II (\cf~Refs.~\cite{Sinha:2001zz, Fischer:2003zz}), but we wish to stress again that we treat the rotational component as a small perturbation. As thoroughly explained in Sec.~\ref{sec:Euler}, the origin of the small ``rotational'' component arises from a relativistic coordinate transformation performed from a Lagrangian to a Eulerian\./\.observer's frame. Actually, the fluid flow is still irrotational, but it is the fluid's space-time which is dragged w.r.t.~the Eulerian frame. We analyse the properties of the Eulerian metric, which are somewhat similar to the one of the Kerr metric, but we also refer to App.~\ref{DerivationADM} for calculational details about its derivation. Equipped with the Eulerian metric, we solve its Klein-Gordon equation in Sec.~\ref{sec:Klein-Gordon}. Then, in Sec.~\ref{sec:Super-Radiance-in-the-Frequency-Domain}, we derive the conditions for the occurence of superradiance, and finally give a Summary and Outlook in~\ref{sec:Summary-and-Outlook}.

\section{System and Equations of Motion}
\label{sec:System-and-Equations-of-Motion}

Our starting point is a slowly-rotating $d$-dimensional Bose-Einstein condensate in its own gravitational potential, being described by a complex scalar field $\hat{\psi}$ with Grand-Canonical Hamiltonian
\begin{align}
\begin{split}
	\hat{\Hcal}
		&=
								\int\limits_{V}\!\d^{d}x\.
								\Bigg\{
									\hat{\psi}^{\dagger}
									\bigg[
										- \frac{1}{2m}\triangle
										+
										\mu
										+
										{\rm i}\,\fett{\Omega} \cdot
										\big(
											\fett{x} \times \!\fett{\nabla}
										\big)
									\bigg]
									\hat{\psi}
									\\[1mm]
		& \,\quad
									+
									\frac{ 1 }{ 2 }
									\int\limits_{V}\!\d^{d}y\;
									\hat{\psi}^{\dagger}( x )\.\hat{\psi}^{\dagger}( y )
									\frac{C}{| x -y |^{d - 2}}_{}\.
									\hat{\psi}( y )\.\hat{\psi}( x )
								\Bigg\}
								\; ,
								\label{eq:hamiltonian}
\end{split}
\end{align}
where $\mu$ is the chemical potential and $V$ is the spatial volume of the halo. The third term on the right-hand side is due to a rotation around the axis $\fett{\Omega}$. The condensate is supposed to be constituted by a number of $N \gg 1$ particles, and self-bound by a gravitational potential, with effective (attractive) interaction strength $C > 0$. We set the Planck constant $\hslash$ equal to one, and express energy in units of the parameter $m$.

In Bose-Einstein condensates, the quantum state $\hat{\psi}$ consists of two components: a highly-occupied ground state, condensate part $\Psi\!\defas\!\langle | \hat{\psi} | \rangle$, which shall here be described by a classical field\footnote{We can describe the condensate part as a classical field because it is precisely a highly occupied state. This fact becomes clearer when expanding the field $\hat{\psi}$ into creation and annihilation operators, $\hat{a}^{\dagger}$ and $\hat{a}$, respectively. Due to the high occupation number $N \gg 1$, we have for the zero-mode, $\hat{a}_{{\bf k} = 0}\.| N \rangle^{}_{\textsc{bec}} = \hat{a}_{{\bf k} = 0}\.\sqrt{N\.}\.| N - 1 \rangle^{}_{\textsc{bec}} \approx \sqrt{N\.}\.| N \rangle^{}_{\textsc{bec}}$, which tantamounts to replacing $\hat{a}_{{\bf k} = 0}$ by the $c$-number $\sqrt{N\.}$.},
and a quantum-fluctuation part $\hat{\phi}$, \ie, $\hat{\psi} \equiv \Psi + \hat{\phi}$. In the so-called Madelung representation \cite{Madelung-Representation} the condensate part reads
\begin{align}
	\Psi
		&\equiv
								\sqrt{n_0}\,
								\erm^{\irm S}
								\; ,
\end{align}
with $n_{0}$ being the ground-state particle-number density, and the phase $S$ is the potential of the longitudinal part of the velocity.

We thus obtain a modified Gross-Pitaevskii equation (\cf, \eg, \cite{Sinha:2001zz,Fischer:2003zz, RindlerDaller:2011kx})
\begin{align}
	\irm\,\partial_{t} \psi
		&=
								\left[
									-
									\frac{1}{2m} \Delta
									+
									\Xi
									+
									{\rm i}\,\fett{\Omega} \cdot
									\big(
										\fett{x} \times \!\fett{\nabla}
									\big)
								\right]
								\psi
								\, ,
								\label{eq:EOM}
\end{align}
where $\psi$ is the normalised wave function of the condensate, $\Xi$ denotes the non-local gravitational potential induced through the condensates density $\rho \defas \psi^{\dagger} \psi$. In particular, the potential satisfies Poisson's equation
\begin{align}
	\Delta \Xi
		&=
								-\,
								\Omega_{d}\.C\.\rho
								\, , 
\end{align}
where $\Omega_{d} \defas 2 \pi^{d} / \Gamma( d / 2 )$, and the latter is the gamma function. We obtain two independent real equations
\vs{-3mm}
\begin{subequations}
\begin{align}
	\partial_{t} n_{0}
	+
	\fett{\nabla} \cdot \fett{j}
		&=
								\fett{\Omega} \cdot
								\big(
									\fett{x} \times \!\fett{\nabla} n_{0}
								\big)
								\, ,
								\label{eq:one}
								\\[2mm]
	\partial_{t} S
	-
	\frac{ 1 }{ 2 m }\.\frac{ 1 }{ \sqrt{n_{0}\.} }&\.\Delta \sqrt{n_{0}\.}
	+
	\frac{ m }{ 2 }
	\big(
		\fett{\nabla} S
	\big)^{2}
	+
	\mu
	+
	\Xi \nonumber \\
		&=
								m\,\fett{\Omega} \cdot
								\big(
									\fett{x} \times \!\fett{\nabla} S
								\big)
								\, , 
								\label{eq:two}
	\qqqq
\end{align}
\end{subequations}
with $\fett{j} \defas n_{0} \fett{\nabla} S$. The second term in Eq.~\eqref{eq:two} is called the {quantum potential}. It has dimension of energy, and it can be shown to be related to the trace of an intrinsically quantum-induced stress-energy tensor (\cf~Ref.~\cite{Barcelo:2005fc}). In the following we will come back to this quantity, and derive a condition under which it can be neglected.

\section{Stability Analysis}
\label{sec:Stability-Analysis}

To this end we expand the background particle-number density as well as the background phase up to first order, and furthermore treat the rotation terms as first-order perturbations. In doing so we allow all perturbations to have a general coordinate dependence. This is important, because the system might be stable under perturbations that respect its symmetry, and unstable with regard to those that do not. We expand
\begin{subequations}
\begin{align}
	n( \fett{x} , t )
 	 	&=
								n_{0}( r )
								+
								\delta n( \fett{x} , t )
								+
								\text{higher orders} \,,
								\label{eq:nexpand}
								\\[2mm]
	S( \fett{x} , t )
		&=
								S_{0}( r )
								+
								\delta S( \fett{x} , t )
								+
								\text{higher orders}
								\, ,
								\label{eq:sexpand}
								\\[2mm]
	\fett{\Omega}(\fett{x}, t )
	&=
								\delta \fett{\Omega}( \fett{x},t )
								+
								\text{higher orders}
								\,.
								\label{eq:Omegaexpand}
\end{align}
\end{subequations}
Note that we treat $\fett{\Omega}$ as a small perturbation. This choice is not only important to stabilise our results, but we shall show that it is also physically motivated.

\subsection{Background}
\label{sec:Background}

\begin{figure}[t]
\centering
	\includegraphics[width=0.9\linewidth]{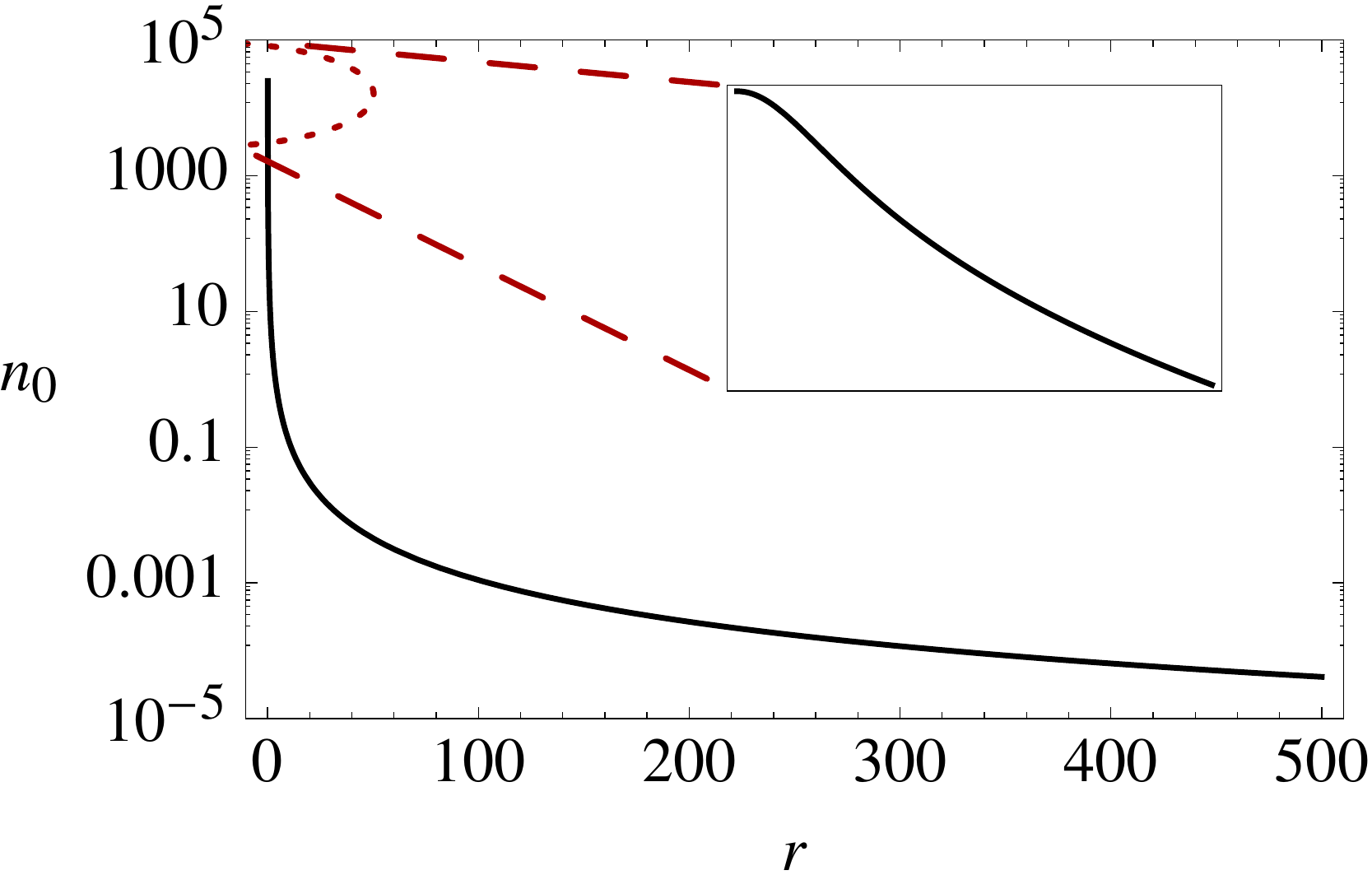}\\
	\includegraphics[width=0.9\linewidth]{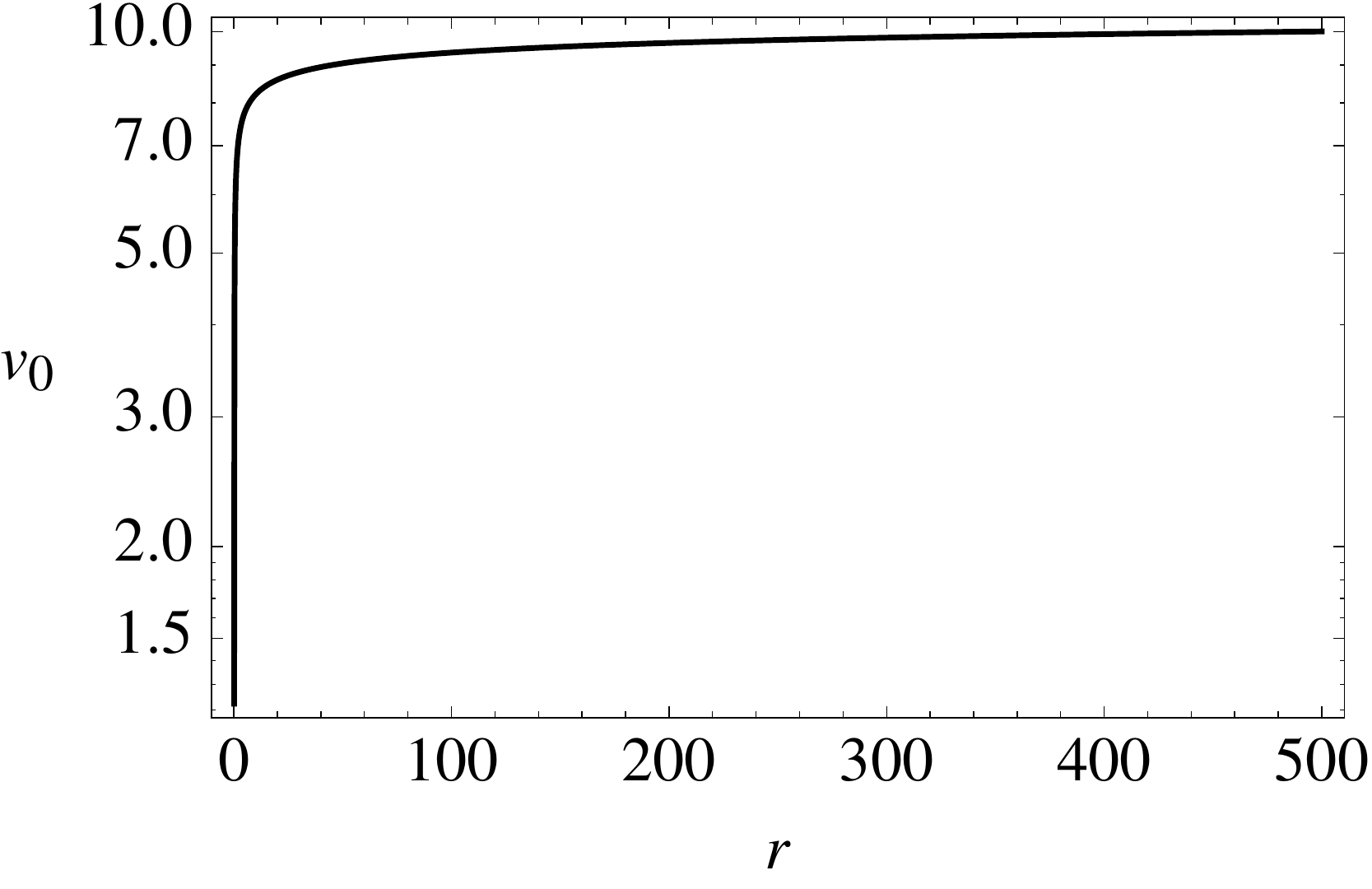}
	\caption{Particle-number density $n_{0}( r )$ (left) and velocity $|\fett{v}_{0}( r )|$ (right) for $d = 3$
		in units where $m = 1$, and for $r_{\text{max}} = 500$ as well as $|\fett{v}_{0}( r_{\text{max}} )| = 10$.}
	\label{fig:v&n}
\end{figure}

Here we consider the static case at the level of the background perturbations, and then add the perturbations in the stability analysis (see the following section). Equation~\eqref{eq:one} is at the 
background level
\begin{align}
 	\fett{\nabla} \cdot \left[ n_{0}\;\!\fett{\nabla} S_{0} \right]
		&\simeq
								0
								\, ,
\end{align}
which yields the solution of the background fluid flow
\begin{align}
 \label{velocitypotential}
	\!\fett{\nabla} S_{0} ( r )
		&=
								\frac{\fett{v}_{00}}{r^{d-1}} \frac{1}{n_{0}( r )} \asdef \fett{v}_{0}
								\, .
\end{align}
Plugging this solution into Eq.~\eqref{eq:two} and neglecting the quantum potential we obtain an approximate expression for the chemical potential,
\begin{align}
	\mu
		&\simeq
								\frac{ 1 }{ 2 } m \fett{v}_{00}^{2} \frac{1}{r^{2d-2}} \frac{1}{n_{0}^{2}( r )}
								+
								C
								\int\!\d^d y\;
								\frac{n_{0}(|\fett{y}|)}
								{| \fett{x} - \fett{y} |^{d-2}}
								\, ,
								\label{eq:chemicalPot}
\end{align}
which we assume to be spatially constant, for simplicity. Then we obtain from the Laplacian of the above equation,
\vs{-2mm}
\begin{align}
	\bar{n}_{0}( r )
		&\simeq
								\frac{ 1 }{ r^{d-1} } \partial_{r} r^{d-1} \partial_{r}
								\bigg[
									\frac{ 1 }{ r^{2d - 2} }
									\frac{ 1 }{ \bar{n}_{0}^{2}( r ) }
								\bigg]
								\, , \label{eq:no}
\end{align}
where $\partial_{r}$ is the radial derivative in polar coordinates, and we have defined
\begin{align}
	\bar{n}_{0}( r )
		&\defas
								\frac{ 1 }{ \sqrt[3]{ \bar{c}\.}\.}\,n_{0}( r )
								\, ,
	\qquad
	\bar{c}
		\defas
								\frac{ 1 }{ 2 }\.\bar{m}\.\fett{v}_{00}^{2}
								\, ,
	\qquad
	\bar{m}
		\defas
								\frac{ m }{ \Omega_{d}\.C }
								\, .
\end{align}
The asymptotic behaviour of the background solution is to a good approximation \cite{Kuhnel:2013uxa}
\begin{subequations}
\begin{align}
	n_{0}( r )
		&\xrightarrow{\,r\.\ra\.\infty\,}
								\;\sim
								1 / { r^{d - 1} }
								\; ,
								\label{eq:n0-large-r}
								\\[3mm]
	\!\fett{v}_{0}( r )
		&\xrightarrow{\,r\.\ra\.\infty\,}
								\textbf{const}
								\, .
								\label{eq:v0-large-r}
\end{align}
\end{subequations}
To get a better intuition of the above, we solve numerically for $n_{0}( r )$ and $v_{0}( r )$ in the case of $d = 3$, see Figs.~\ref{fig:v&n}. We have checked that the general behaviour of $n_{0}( r )$ and $v_{0}( r )$ is essentially 
independent of the chosen initial conditions, which have to be imposed to Eqs.~(\ref{eq:chemicalPot}-\ref{eq:no}). 

The above solutions for the background density and velocity have been derived under the assumption that the quantum pressure can be neglected.\footnote{This is commonly called the Thomas-Fermi approximation \cite{Dalfovo:1999zz,Boehmer:2007um,Sikivie:2009qn}.} More precisely, validity requires
\begin{align}
	\frac{ 1 }{ 2 m }\.\frac{ 1 }{ \sqrt{n_{0}\.} }\.\Delta \sqrt{n_{0}\.}
		&\simeq
								0
								\, .
								\label{eq:neglecting-the-quantum-pressure}
\end{align}
Now, in the outer region of a sufficiently large halo (in which we will be mainly interested) we have that the quantum-pressure term actually diminishes approximately as $\sim r^{-2}$. This means that the quantity $x \defas 1 / ( m\.r_{\text{max}}^{2} )$ has to be sufficiently small. Eq.~\eqref{eq:chemicalPot} suggests that $x$, which has the dimension of energy, has to be compared to the chemical potential $\mu$. Thus, we get the approximate condition
\vs{-2mm}
\begin{align}
	\mu / x
		&=
								\mu\.m\.r_{\text{max}}^{2}
		\gg
								1
								\, ,
								\label{eq:condition-for-neglecting-the-quantum-pressure}
\end{align}
which translates to the condition for the maximal extent of the halo, \ie, $r_{\text{max}} \gg {\mu\.m\.}^{-1/2}$.

\subsection{Stability}
\label{sec:Stability}

We now perform a stability analysis. Using Ansatz~(\ref{eq:nexpand})--(\ref{eq:Omegaexpand}), and the zeroth-order results for the velocity and density, Eqs.\,(\ref{velocitypotential})--(\ref{eq:no}), we obtain up to first order
\begin{subequations}
\begin{align}
	\partial_{t}\delta n( \fett{x} , t )
		&\simeq
								-\.
								\fett{\nabla}\!\cdot\!
								\bigg[
									\!\fett{\nabla} S_0\,
									\delta n
					+
									\frac{ 1 }{ m }\.
									n_{0}\,
									\fett{\nabla}\delta S
								\bigg]
								 \nonumber \\
			&\qquad \qquad	+
								\fett{\Omega} \cdot
								\big[
									\fett{x} \times \!\fett{\nabla} \delta n
								\big]
								\, ,
								\label{eq:deltan}
								\displaybreak[1]
								\\[3mm]
\begin{split}
	\partial_{t}\delta S( \fett{x} , t )
		&\simeq
								-\,
								\fett{\nabla} S_0
								\cdot
								\fett{\nabla} \delta S
								+
								\fett{\Omega} \cdot 
								\big[
									\fett{x} \times \!\fett{\nabla} \delta S
								\big] \\
 	& \qquad \qquad -
								\int\!\d^{d}y\,
								\frac{C}{| y - x |^{d - 2}}\.
								\delta n( y , t )
								\, ,
								\label{eq:deltas-1}
\end{split}
\end{align}
\end{subequations}
where we have suppressed some dependences for notational simplicity. In particular, $\fett{\Omega}$ can vary with space and time. Acting with $\triangle$ on Eq.~\eqref{eq:deltas-1}, we find
\begin{align}
	\delta n
		&\simeq
								\frac{ 1 }{ \Omega_{d}\.C }\,
								\triangle
								\bigg(
									\partial_{t}\delta S
									+
									\!\fett{\nabla} S_0
									\cdot
									\fett{\nabla} \delta S
									-
									\fett{\Omega} \cdot 
									\big[
										\fett{x} \times \!\fett{\nabla} \delta S
									\big]
								\bigg)
								\. .
								\label{eq:delta-n-in-terms-of-delta-S}
\end{align}
Plugging Eq.~\eqref{eq:delta-n-in-terms-of-delta-S} into Eq.~\eqref{eq:deltan} yields 
\begin{align}
	& \triangle \bigg( \partial^{2}_{t}\delta S + \!\fett{\nabla} S_0 \cdot \fett{\nabla} \partial_{t} \delta S 
	 - \fett{\Omega} \cdot \big[ \fett{x} \times \!\fett{\nabla} \partial_{t} \delta S \big] \bigg) \notag \\ &\quad
\simeq -\. \fett{\nabla}\!\cdot\! \Bigg[\frac{ \Omega_{d}\.C }{ m }\. n_{0}\, \fett{\nabla}\delta S 
 + \!\fett{\nabla} S_0\; \triangle \bigg( \partial_{t}\delta S \bigg. \nonumber \\
&\qquad\qquad \quad\bigg. + \!\fett{\nabla} S_0 \cdot \fett{\nabla} \delta S - \fett{\Omega} \cdot \big[ \fett{x} \times \!\fett{\nabla} \partial_{t} \delta S \big] \bigg) \Bigg] \. . \vphantom{\Bigg|}
\end{align}
Now, taking into account Eqs.~(\ref{eq:n0-large-r},b), we are for large $r$ approximately left with
\begin{subequations}
\begin{align}
	&\Big[
		\partial_{t}
		+
		\!\fett{\nabla} S_0(r) \!\cdot\!\fett{\nabla}
		-
		\fett{\Omega} (\fett{x},t)\cdot 
		\big(
			\fett{x} \times \!\fett{\nabla}
		\big)
	\Big]^{2}\.
	\delta \tilde{S} (\fett{x},t)
		\simeq
								0
								\,,
								\label{eq:Xi-Eq-large-r}
\intertext{where we have restored all dependences. The quantity $\delta \tilde{S}( \fett{x} , t ) \defas \triangle\delta S( \fett{x} , t )$ can be thought of as the gradient of the velocity perturbation. Hence, any pulse keeps its shape at large distance from the centre of the halo. Along similar lines, Eq.~\eqref{eq:deltan} becomes}
	&\Big[
		\partial_{t}
		+
		\!\fett{\nabla} S_0( r )\!\cdot\!\fett{\nabla}
		-
		\fett{\Omega}( \fett{x} , t ) \cdot 
		\big(
			\fett{x} \times \!\fett{\nabla}
		\big)
	\Big]\.
	\delta n( \fett{x} , t )
		\simeq
								0
								\, ,
								\label{eq:Xi-Eq-large-r2}
\end{align}
\end{subequations}
Thus, our stability result also holds for arbitrary perturbations in the particle-number density. Moreover, these findings are true for all dimensions greater than two.

\section{Eulerian Metric}
\label{sec:Euler}

As mentioned above we treat the rotation as a small perturbation \wrt~the longitudinal velocity component. Here we include such a suppressed rotational velocity component by performing a gauge transformation of an irrotational fluid flow in a synchronous\./\.comoving coordinate system to an observer's coordinate system. We call the synchronous/comoving coordinate systems to be the Lagrangian frame; we call the observer's coordinate system the Eulerian frame. 

The essential idea of such a coordinate/gauge transformation is, 
that its intrinsic non-linear nature induces naturally a small amount of frame dragging \cite{Rampf:2013dxa}. Thus, even that the fluid motion is irrotational, the fluid's Lagrangian space-time is dragged \wrt~the Eulerian frame. Such a frame dragging is essential to observe superradiance.

For simplicity we set $d = 3$ from now on. We transform such an irrotational fluid flow to the Eulerian frame with the following line element
(see App.~\ref{DerivationADM})
\vs{-1mm}
\begin{align}
 \begin{split}
	\d s^{2} = &- \left( 1 - \fett{v}_{\rm L}^{2} \right) \,\d t^{2} -2 ( \fett{v}_{\rm L} + \fett{v}_{\rm T} ) \!\cdot\!\d \fett{x}\,\d t \\
	&\qquad	 + \left( 1 - \frac 5 2 \fett{v}_{\rm L}^{2} \right) \d \fett{x}^{2} \,,	\label{ADM}
 \end{split}
\end{align}
where $\fett{v}_{\rm L} \defas \fett{\nabla} S_{0}$ and $\fett{v}_{\rm T} \defas \fett{\nabla} \times \fett{T}$ denote the longitudinal and a perturbatively suppressed transverse fluid velocity, respectively; $S_{0}$ is the background velocity potential, see Eq.\,(\ref{velocitypotential}). Calculational details about the transverse field $\fett{T}$ can be found in the App.~\ref{DerivationADM}, but we do not need its explicit form in the following, since we shall assume a constant amplitude in the latter, for simplicity. In the above line element we have set the local speed of sound to one, i.e., $c=1$. This approximation also means that we neglect pressure perturbations; this is consistent with the neglect of the quantum pressure term (see Sec.~\ref{sec:Background}), but including pressure perturbations is far beyond the scope of this paper. We leave this issue for a forthcoming paper.

In this paper we assume that the longitudinal velocity component is much larger than the transverse component
\vs{-2mm}
\be \label{smallness}
	\frac{\left|\fett{v}_{\rm T}\right|}{\left|\fett{v}_{\rm L}\right|} \ll 1 \,.
\ee
It is difficult to estimate the precise amount of the suppression of the transverse velocity, since we heavily rely on a Lagrangian extrapolation far into the non-linear regime. On cosmological scales, where the solutions are formally valid, we expect that the transverse velocity should be suppressed by a factor $\sim 10^{-3} - 10^{-5}$.
Note also that we can discard the term $\propto \fett{v}_{\rm L}^{2}$ in the space-space component in Eq.~\eqref{ADM} for sufficient small enough velocities.\footnote{Actually, the very term in the space-space component in Eq.\,(\ref{ADM}) is of ${\cal O}\!\left( \fett{v}_{\rm L}^4 \right)$ as can be seen when calculating the proper time between two events along the worldline \cite{Barcelo:2005fc,Rampf:2013dxa}.}

Generally, the above line element is valid for an arbitrary geometry. In the following we consider an approximate spherical symmetry and also assume a stationary and convergent fluid flow \cite{Unruh:1980cg}. 
Now, performing a temporal gauge transformation according to
\begin{align}
	\tau( t, r ) &=	t+ \int^{r}\! \d r'\;	\frac{ v_{r}( r') }{ 1 - v_{r}^2(r') } \,,
\label{eq:tau-transformation}
\end{align}
and we set $\fett{v}_{\rm T} \defas {\hat {\fett{e}}}_{\phi} v_{\rm T}$, 
 the metric~(\ref{ADM}) becomes 
\begin{widetext}
\begin{align}
	\d s^{2} = &- \big( 1 - v_{r}^{2} \big)\. \d \tau^{2} + \frac{1}{1 - v_{r}^{2}}\. \d r^{2} + r^{2}\. \d \phi^{2}	 
		- 2\. r\.v_{\rm T}\, \d \phi \, \d \tau + 2\.r \frac{v_{r}\.v_{\rm T}}{1 
									-	v_{r}^{2}}\,	\d \phi \,\d r \,, \label{eq:metricIrrotational}
\intertext{or, equivalently,} 
	\d s^2 =	&- \Big[ 1 - \left( v_{r}^{2} + v_{\rm T}^2 \right) \Big]\. \d \tau^{2} + \frac{1}{1 - v_{r}^{2}}\. \d r^{2}	+	r^2 \left(	\d \phi - \frac{v_{\rm T}}{r} \d \tau \right)^2 
		 +	2\. r \, \frac{v_{r}\.v_{\rm T}}{1 - v_{r}^{2}}\, \d \phi \,\d r \,. \label{eq:metricIrrotationalA}
\end{align}
\end{widetext}
\newpage
Again, we have $v_{\rm T}/v_r \ll 1$, in accordance with the requirement~(\ref{smallness}). Note also that we have restricted the line element to the equatorial plane (i.e., $\dd \theta = 0$).
Some properties of the above metric are identical to the one of a Kerr metric, but differs by one (i.e., the last) property:
\begin{itemize}
	\item It is stationary, i.e., it does not explicitly depend on time. 
	\item It is axisymmetric, i.e., it does not depend explicitly on $\phi$.
	\item It is not static, i.e., 
		it is not invariant under time reversal $\tau \rightarrow - \tau $.
	\item It reduces to a Schwarzschild like metric in the limit $v_{\rm T} \rightarrow 0$, if the fluid velocity smoothly exceeds the speed of sound $v_{r} \defas 1 - \alpha (r - R) - {\cal O}(\alpha^2)$. This also means that we expect the occurence of Hawking radiation	\cite{Unruh:1980cg,Dvali:2012en} (see also \cite{Kuhnel:2013uxa}).
	\item There is a coordinate singularity when $v_{r} \rightarrow 1$ (i.e., when it approaches the local speed of sound), and a curvature singularity for $r \rightarrow 0$. Indeed, since $\lim_{r \rightarrow 0}v_{r} \propto 1/r^2$,
 the 4-Ricci curvature is $\lim_{r \rightarrow 0}{}^{(4)}R \propto 1/[r^6 (1+v_{\rm T}^2)] = \infty$.
	\item The metric is invariant under the simultaneous inversion: $\tau \rightarrow - \tau$, $\phi \rightarrow - \phi$.\footnote{Note that the last term in~(\ref{eq:tau-transformation}) also flips the sign under the inversion $\tau \rightarrow - \tau$, as it should.}
	\item The metric is not asymptotically flat, i.e., the metric does not reduce to the Minkowski metric in the limit $r \rightarrow \infty$: 
          $\dd s^2 = -C^2 \d \tau^{2} + C^{-2} \d r^{2} + r^{2} \d \phi^{2}	 
		- 2 r v_{\rm T}\, \d \phi \, \d \tau + 2 r v_{\rm T} D \d \phi \d r$, with $C= 1-v_r^2= const$, and $D= v_r C^{-2} = const$.
\end{itemize}
Thus, because of the last property, the metric~(\ref{eq:metricIrrotationalA}) is not identical to the Kerr metric. 
The above metric is not asymptotically flat due to the fact, that the frame dragging is in principle apparent even at radial infinity, despite the fact that also the background density of the condensate is vanishing at infinity (see Fig.~(\ref{fig:v&n})). 
This feature of the Eulerian metric is the result of the change of inertial frames (Mach's principle) between the condensate and the observer. The occurence of superradiance is not affected by the fact that our metric~(\ref{eq:metricIrrotationalA}) is not asymptotically flat.


\section{Klein-Gordon Equation with Tortoise Coordinates}
\label{sec:Klein-Gordon}

Here we solve the relativistic Klein-Gordon equation for the linearised velocity-perturbation potential $\hat{\Phi} \equiv 1 / ( 2\.m\.\irm\.\sqrt{n_{0}\.} )\.\big( \text{e}^{-\irm\.S}\.\hat{\phi} + \text{e}^{\irm\.S}\.\hat{\phi}^{\dagger} \big)$, which reads (\cf~Refs.~\cite{Unruh:1980cg, Barcelo:2005fc})
\be \label{KGgeneral}
	\partial_{\mu} \left( \sqrt{-g} g^{\mu\nu} \partial_\nu \right) \hat{\Phi} = 0 \,,
\ee 
with $g = \det g_{\mu\nu}$, and summation over repeated space-time indices $\mu, \nu$, is implied. As mentioned earlier, we set $d = 3$ for simplicity, although our calculations may be easily generalized to higher dimensions.

Note that we make use of two essential simplifications at this step. First, we suppress the ``spatial dimension'' related with the altitudinal angle $\dd \theta$. Second, we assume that $v_{\rm T}$ is constant. Relaxing these approximations do not lead to any conceptual difficulties, but would imply a dramatical increase in computational power. In this paper, we use these simplifying approximations mainly since we are only interested in the qualitative behaviour of superradiance, and we leave computational more demanding approaches for future work. Note however, that ${v_{\rm T}}$ should be generally time dependent, since we expect that the negative momentum transfer due to superradiance leads to a decreasing ${v_{\rm T}}$ in time. However, considering the typically large amount of rotational energy as compared to that of the small amount of energy of a single scattered wave, it is a good approximation to neglect the time-dependence of ${v_{\rm T}}$ for each individual scattered wave, which we do so in the following for simplicity.

Now we turn back to the Klein-Gordon Eq.~\eqref{KGgeneral}. Let us therefore decompose the quantum field $\hat{\Phi}$ into creation and annihilation operators, $\hat{\alpha}^{\dagger}$ and $\hat{\alpha}$, respectively, \ie, $\hat{\Phi} \equiv \sum \hat{\alpha}\.f + \hc$, for some mode function $f$. Then, using our result in Eq.~\eqref{eq:metricIrrotational}, the Klein-Gordon equation becomes \cite{Berti:2004ju}
\begin{widetext}
\begin{align} 
	&	r^{2} \big[ 1+v_{\rm T}^{2} -v_{r^{*}}^{2} \big] \Delta_1^{2} f_{,r^{*}r^{*}} - 2 r v_{\rm T} v_{r^{*}} \Delta_1 f_{,r^{*} \phi} + 2 r \Big[ 1 +v_{\rm T}^{2} - v_{r^{*}} \big\{ v_{r^{*}} + r \Delta_1 v_{r^{*},r^{*}}\big\} \Big] \Delta_1 f_{,r^{*}} 
		 \nonumber
		 \displaybreak[1] \\[1mm]
 &+\frac{r v_{\rm T}^{2}}{1 -v_{r^{*}}^{2}} \Bigg( \bigg[ 2 v_{r^{*}} + r \frac{1 + v_{r^{*}}^{2}}{1-v_{r^{*}}^{2}} \Delta_1 v_{r^{*},r^{*}} \bigg] f_{,t} + r v_{r^{*}} \Delta_1 f_{,tr^{*}} \Bigg) 
 - \frac{r^{2} \big(1 - \big[ 1 + v_{\rm T}^{2} \big] v_{r^{*}}^{2} \big)}{\big( 1 -v_{r^{*}}^{2} \big)^{2}} f_{,tt} \nonumber\displaybreak[1] \\[1mm]
 &+ f_{,\phi \phi} - v_{\rm T} \left( v_{r^{*}} + r \Delta_1 v_{r^{*},r^{*}} \right) f_{,\phi} - 2 r \frac{v_{\rm T}}{1 -v_{r^{*}}^{2}}	f_{,t \phi}	+ \frac{r^{2} v_{\rm T}^{2}	v_{r^{*}}}{1-v_{r^{*}}^{2}} \Delta_1 f_{,tr^{*}} = 0 \,,
\label{eq:KG1}
\end{align}
\end{widetext}
where we have introduced the tortoise coordinates $r^{*}$ which are
\be \label{tortoise}
	\frac{\dd r^{*}}{\dd r^{\.\.\.}} \asdef \Delta_1 \,, \qquad \frac{\dd}{\dd r} = {\Delta_1} \frac{\dd}{\dd	r^{*}} \,, \qquad \Delta_1 =	\frac{1}{1-{v_{r^{*}}^{2}}} \,.
\ee
$\Delta_1$ was derived from the requirement that the purely spatial and purely temporal parts of the metric~(\ref{eq:metricIrrotational}) are conformally invariant, \ie, 
\vs{-1mm}
\be
	X\.\big( - \dd \tau^{2} + \d {r^{*}}^{2} \big) \stackrel{!}{=} - \big(1- {v_{r^{*}}^{2}} \big) \dd \tau^{2} + \frac{1}{1-{v_{r^{*}}^{2}}} \dd r^{2} \,,
\ee
where $X$ is found to be $X = 1 / \Delta_{1}$. 

To solve the Klein-Gordon equation~(\ref{eq:KG1}), we fix $v_{\rm T}$ to be a small\footnote{$v_{\rm T}$ has to be small because of the requirement from~(\ref{smallness}), which itself is based on the App.~\ref{DerivationADM}.} constant, and we use the background solution of the longitudinal velocity~$v( r ) \simeq v_{0}( r )$, which is also depicted in the right panel of Fig.~(\ref{fig:v&n}). Since the dependence of $\Delta_1$ (through $v_r$) on the tortoise coordinate $r^{*}$ is non-trivial, we numerically integrate 
\vs{-1mm}
\be
	 r^{*}( r ) \equiv \int {\dd r^{*}} = \int \Delta_1 {\dd r} \,,
\ee
and then use this result explicitly in the Klein-Gordon equation~(\ref{eq:KG1}).

\section{Superradiance in the Frequency Domain}
\label{sec:Super-Radiance-in-the-Frequency-Domain}

Using the tortoise coordinates from the previous section, and utilising a decomposition of the mode function $f$ into base elements
\begin{align}
	f( r^{*},\.m,\.\omega )
		&=
								z_{\omega,\.m}( r^{*} )\.
								\varphi_{\omega,\.m}( r^{*} )\.
								\text{e}^{- \irm m \phi}\.
								\text{e}^{\irm \omega t}
								\, ,
								\label{eq:f-base-element}
\end{align}
we find, after plugging the above expression into Eq.~\eqref{eq:KG1}, a second-order differential equation equation of the structure
\begin{align}
	\Big[
		\alpha_{2}
 \.\partial^{2}_{r^{*}}
		+
		\alpha_{1}
 \.\partial_{r^{*}}
		+
		\alpha_{0}
	\Big]
	\Big[
		z_{\omega,\.m}( r^{*} )\.\varphi_{\omega,\.m}( r^{*} )
	\Big]
		&=
								0
								\; ,
								\label{eq:second-order-equation-z-times-varphi}
\end{align}
where $\alpha_i \equiv \alpha_i(\omega,\.m)$.
We determine the function $z_{\omega,\.m}$ such that the coefficient of $\partial_{r^{*}}\varphi_{\omega,\.m}$ vanishes (\cf~Ref.~\cite{Berti:2004ju}). Then, the quasi-normal form of the differential equation for $\varphi_{\omega,\.m}$ reads
\begin{align}
	\partial^{2}_{r^{*}} \varphi_{\omega,\.m}( r^{*} )
	=-
	\Big[
		\Vrm( r^{*},\.\omega,\.m )
		+
		\irm\.\Gamma( r^{*},\.\omega,\.m )
	\Big]
	\varphi_{\omega,\.m}( r^{*} )
							.
								\label{eq:V-and-rstar-times-Gamma-large-rstar}
\end{align}
Here, both $\Gamma( r^{*},\.\omega,\.m )$ and $\Vrm( r^{*},\.\omega,\.m )$ are real functions, and are given by
\begin{subequations}
\begin{align}
	\Gamma( r^{*},\.\omega,\.m )
		&=
								-
								\frac{m\.v_{\rm T}\.v_{r^{*}}^{3}
									\big(
										1
										-
										v_{r^{*}}^{2}
									\big)
								}{
									{r^{*}}^2
									\big(
										1
										-
										v_{r^{*}}^{2}
										+
										v_{\rm T}^{2}
									\big)
								}
								\; ,
								\label{eq:Gamma}
\end{align}
where we introduced the shorthand notation $v_{r^{*}} \equiv v_{0}(r^*)$, and we shall indicate spatial derivatives w.r.t.~$r^*$ with a prime in the following, and
\begin{widetext}
\begin{align}
	&\Vrm( r^{*},\.\omega,\.m )
		=
								\frac{ 1 }
								{
								\Big(
										{r^{*}}^{2}
										\big[
											1
											-
											v_{r^{*}}^{2}
										\big]^{2}
										\big[
											1
											+
											v_{\rm T}^{2}
											-
											v_{r^{*}}^{2}
										\big]^{2}
								\Big) }
								\Bigg[
								r^{*} v_{r^{*}}^3
								\Big(
										r^{*} v_{\rm T}^{2}
										\big[
											2 + v_{\rm T}^{2}
										\big]
										v_{r^{*}}''
										-
										2
										\big[
											4
											+
											3 v_{\rm T}^{2}
										\big]
										v_{r^{*}}' 
								\Big)
								\notag
								\displaybreak[1]
								\\[2mm]
								&\.
								+
								\bigg\{
										v_{\rm T}^{2}
										\big(
											{r^{*}}^2 \omega^{2}
											-
											2 m r^{*} \omega v_{\rm T}
											-
											v_{\rm T}^{2}
										\big)
										-
										5
										\big[
											2
											+
											m^{2}
										\big]
										+ 
										\Big(
											{r^{*}}^2 \omega^{2}
											-
											2 m r^{*} \omega v_{\rm T}
											-
											\big[
												8
												+
												5 m^{2}
											\big]
											v_{\rm T}^{2}
										\Big)
								\bigg\}
								v_{r^{*}}^8
								\notag
								\displaybreak[1]
								\\[2mm]
\begin{split}
								&\.
								+
								\Big(
									\big[
										5
										+
										m^{2}
									\big]
									+
									\big[
										2
										+
										m^{2}
									\big]
									v_{\rm T}^{2}
								\Big)
								v_{r^{*}}^{10}
								-
								v_{r^{*}}^{12}
								-
								2 r^{*}
								\big[
									4
									+
									v_{\rm T}^{2}
								\big]
								v_{r^{*}}^{7} v_{r^{*}}'
								+
								2 r^{*} v_{r^{*}}^9 v_{r^{*}}'
								\\[2mm]
								&\.
								+
								\big[
									1
									+
									v_{\rm T}^{2}
								\big]
								v_{r^{*}}^{2}
								\Big\{
										\big(
											1
											+
											5 m^{2}
											-
											4 {r^{*}}^2 \omega^{2}
											+
											8 m r^{*} \omega v_{\rm T}
											+
											v_{\rm T}^{2}
										\big)
										-
										2 {r^{*}}^2 v_{\rm T}^{2} v_{r^{*}}'^{2}
								\Big\}
								\\[2mm]
								&\.
								-
								\big[
										1 + v_{\rm T}^{2}
								\big]
								\bigg\{
										\Big(
											 m^{2}
											 -
											 r^{*} \omega
											\big[
											 	r^{*} \omega
												-
												2 m v_{\rm T}
											\big]
										\Big)
										+
										{r^{*}}^2 v_{\rm T}^{2} v_{r^{*}}'^{2}
								\bigg\}
\end{split}
								\label{eq:V}
								\\[2mm]
								&\.
								+
								 v_{r^{*}}^4
								\bigg\{
										\big[
											1
											+
											v_{\rm T}^{2}
										\big]
										\Big(
											6 {r^{*}}^2 \omega^{2}
											-
											5
											\big[
												1
												+
												2 m^{2}
											\big]
											-
											3 v_{\rm T}
											\big[
												4 m r^{*} \omega
												+
												v_{\rm T}
											\big]
										\Big)
										+
										3 {r^{*}}^2 v_{\rm T}^{2} v_{r^{*}}'^{2}
								\bigg\}
								\notag
								\displaybreak[1]
								\\[2mm]
								&\.
								+
								\!\bigg\{
										10
										\big[
											1
											+
											m^{2}
										\big]
										-
										v_{\rm T}^{2}
										\big(
											4 {r^{*}}^2 \omega^{2}
											-
											8 m r^{*} \omega v_{\rm T}
											-
											3 v_{\rm T}^{2}
										\big)
										-
										2
										\Big(
											2 {r^{*}}^2 \omega^{2}
											-
											4 m r^{*} \omega v_{\rm T}
											-
											\!\big[
												6
												+
												5 m^{2}
											\big]
											v_{\rm T}^{2}
										\Big)\!
								\bigg\} v_{r^{*}}^6
								\notag
								\displaybreak[1]
								\\[2mm]
								&\.
								+
								r^{*} v_{r^{*}}^5
								\Big(
										6
										\big[
											2 
											+
											v_{\rm T}^{2}
										\big]
										v_{r^{*}}'
										-
										r^{*} v_{\rm T}^{2} v_{r^{*}}''
								\Big)
								+
								r
								\big[
										1
										+
										v_{\rm T}^{2}
								\big]
								v_{r^{*}}
								\big(
										2 v_{r^{*}}'
										-
										r^{*} v_{\rm T}^{2} v_{r^{*}}''
								\big)
								\Bigg]
								\. .
								\notag
\end{align}
\end{widetext}
\end{subequations}
\begin{figure}
\centering
	\includegraphics[width=0.9\linewidth]{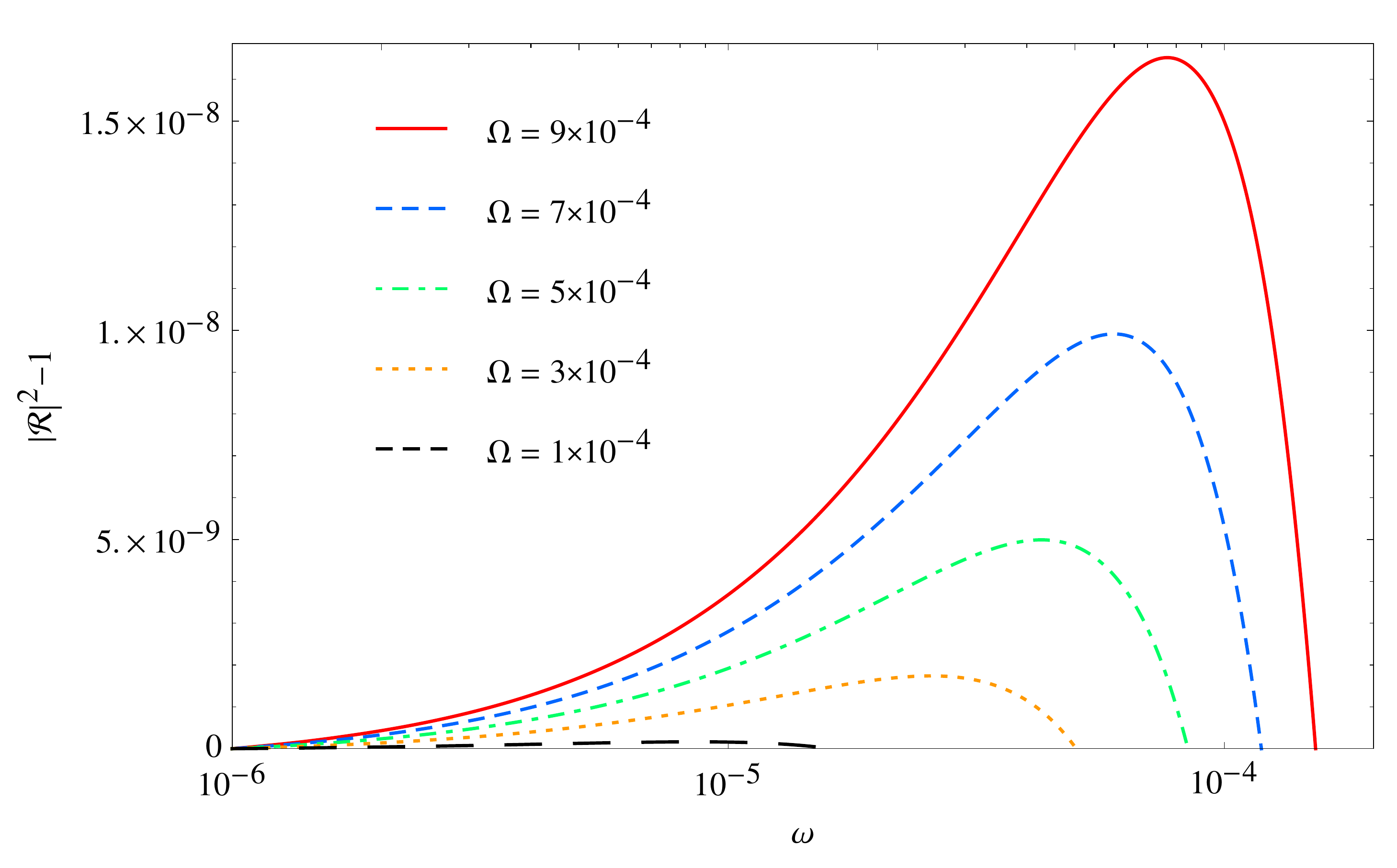}
	\caption{$| \Rcal |^{2} - 1$ as function of frequency for $d = 3$ and
		for various values of ${\Omega} \defas v_{\rm T} / | \fett{v}_{0}( r^{*}_{\text{max}} ) |$.
		The other parameters are $m = 1$, $r^{*}_{\text{max}} = 500$ and $| \fett{v}_{0}( r^{*}_{\text{max}} ) | = 10$.}
\label{fig:Rsquare}
\end{figure}
\!\!Now, as we have seen in Sec.~\ref{sec:Background}, the background velocity approaches a constant for large $r^{*}$. In this limit we find 
\vs{-2mm}
\begin{align}
 	r^{*}\,\Gamma( r^{*},\.\omega,\.m )
		&\simeq
								0
								\; ,
	\qq
	\Vrm( r^{*},\.\omega,\.m )
		\simeq
								\tilde{\omega}^{2}
								\; ,
								\label{eq:Asymptotic-behaviour-of-Gamma-and-V}
\end{align}
where
\vs{-1mm}
\begin{align}
	\tilde{\omega}
		&\simeq
								\omega\,
								\frac{
								\left(
									1
									-
									v_{r^{*}}^2
								\right)
								\sqrt{
										1
										+
										v_{\rm T}^{2}
								\.}}
								{
											1
											-
											v_{r^{*}}^{2}
											+
											v_{\rm T}^{2}
								\.}
								\,
								\; .
								\label{eq:tau}
\end{align}
Note that $\lim^{}_{v_{\rm T}\.\rightarrow\.0} \tilde \omega = \omega$, so the energy of the outgoing wave is identical to the energy of the ingoing wave in this limit. The above relation also means that, for large $r^{*}$, the solution of Eq.~\eqref{eq:V-and-rstar-times-Gamma-large-rstar} for $\varphi \equiv \varphi_{\omega,\.m}$ can be written as
\begin{align}
	\varphi( r^{*} )
		&\simeq
								\text{e}^{- \irm\.\tilde{\omega}\.r^{*}}
								+
								\Rcal\,
								\text{e}^{\irm\.\tilde{\omega}\.r^{*}}
								\, ,
								\qquad r^{*} \rightarrow \infty
								\; ,
								\label{eq:varphi-solution}
\end{align}
which consists of an incident wave of unit amplitude and a scattered one with an amplitude given by the reflection coefficient $\Rcal$. The extraction of (the condensate's rotational) energy by a scattered wave is one example of the phenomenon of {\it superradiance} (\cf~Ref.~\cite{Bekenstein:1998nt} for a review). It occurs if\footnote{This condition can be shown to be equivalent to that of an imaginary Wronskian (\cf~Ref.~\cite{Richartz:2009mi}).}
\vs{-3mm}
\begin{align}
	| \Rcal |^{2}
		&>
								1
								\; .
								\label{eq:R-absolute-square-larger-than-one}
\end{align}
In Fig.~\ref{fig:Rsquare} we show precisely this quantity for the most relevant case of three spatial dimensions, and for various small values of the ratio $\Omega \defas v_{\rm T} / | \fett{v}_{0}( r^{*}_{\text{max}} ) |$ for which our approximation is valid (\cf~Eq.~\eqref{smallness} and also App.~\ref{DerivationADM}). We observe that the system becomes more and more superradiant if this ratio is increased. This can be achieved by either increasing the rotation of the object, or, by decreasing the (outer) radial velocity. However, for all studied parameter values $\Omega$, the corresponding amplification is quite small. 

We find that, as expected, superradiance is only present for small-enough frequencies, \ie, for $\omega < \omega_{\text{max}}$. Here, $\omega_{\text{max}} > 0$ is defined by the condition $| \Rcal( \omega_{\text{max}} ) |^{2} \overset{!}{=} 1$. For the present choice of parameters ($r^{*}_{\text{max}} = 500$ and $| \fett{v}_{0}( r^{*}_{\text{max}} ) | = 10$) we obtain for all studied curves the fixed ``law'' of superradiance which is to a very good approximation 
\vs{-1mm}
\begin{align}
	\omega \,
		&< \,
								0.02\.m\.\Omega
								\; ,
								\label{eq:super-radiance-condition}
\end{align}
which is our central result in this section.

\section{Summary and Outlook}
\label{sec:Summary-and-Outlook}

In this work we have investigated stability and dynamical properties of slowly-rotating gravitationally self-bound Bose-Einstein condensates. Although the model under consideration is rather general, we primarily focussed on respective dark-matter halos. However, we expect that our considerations should also hold for Bose-Einstein condensates of white dwarfs, neutron and boson stars.

First, we derived the modified Gross-Pitaevskii equation and found general solutions for the background particle-number density and proved that the associated background equations of motion are stable in all dimensions. This holds true as long as the rotational component is treated as a small perturbation. We showed that the dynamics of the perturbation of the velocity potential are effectively governed by the Klein-Gordon equation of a newly-derived ``Eulerian metric''.

The latter is based on a relativistic coordinate transformation from the (Lagrangian) fluid frame to an observer's frame, where a small `rotational' component naturally arises because of the inherent non-linearities due to the gravitational instability. Physically, the fluid flow is irrotational but the Eulerian frame is dragged \wrt~the Lagrangian frame, where the fluid is at rest. We thus associate the small rotational component with Lagrangian transverse fields which manifests in a frame dragging. We would like to emphasise that the derived Eulerian metric is generally valid on cosmological scales, and that the reported geometrical correspondence with the one of the Bose-Einstein condensate relies heavily on a Lagrangian extrapolation, \ie, the correspondence holds only approximately.

We then analysed the properties of this Eulerian metric, which shares many properties of the Kerr metric (\eg, the Eulerian metric has an event horizon and an ergosphere, when the fluid velocity becomes transonic), but differs in such that the Eulerian metric is not asymptotically flat. (The latter has no consequence for the occurence of superradiance.)
 
Then, we particularly investigated the effect of superradiance at the vicinity of Bose-Einstein condensates. For convenience, we restricted to a (2+1)-dimensional description \cite{Basak:2002aw,Berti:2004ju}. Our formalism has the advantage that the boundary conditions of the velocity and density are everywhere well-defined and, in particular, our model does not possess a divergent velocity at the origin, as it is the case in the draining bathtub model. We find that superradiance actually does occur, and computed the precise form of the reflection coefficient. We find that, in our general model, amplification takes place when $\omega < \omega_{\text{max}} = 0.02 \.m\. \Omega$ is satisfied, where $\Omega$ is the ratio of transverse and longitudinal velocity of the condensate, and $m$ is the azimutal quantum number of the scattered wave.

As halos typically rotate quite slowly (\cf~Ref.~\cite{Bett:2010}), the energy gain of these waves is naturally rather small. 
Note that such a suppressed rotation is actually a prediction of our model. The precise level of suppression can vary from model to model, so we 
depicted in Fig.~\ref{fig:Rsquare} various values for $\Omega$, ranging from $10^{-3}$--$10^{-4}$. From the results we indeed conclude, that 
 the more rapidly the condensate rotates the more superradiance is expected, and the larger is the maximum frequency below which amplification takes place.
 
Our results may provide an observational tool to discriminate\./\.constrain possible Bose-Einstein condensate dark-matter halo models. In particular, radiation from intense sources such as super-nov{\ae} or gamma-ray bursts in the line of sight behind a possible dark-matter halo will be an optimal test for these class of models. Another set of constrains might come from gravitational lensing, which is expected to be larger then in the conventional cases (\cf, \eg, Ref.~\cite{Dabrowski:1998ac} for Bose stars, and Ref.~\cite{Schunck:2006rk} for scalar-field halos). Of course, similar investigations could be applied to other astrophysical condensates.

\section*{Acknowledgements}

C.R.~acknowledges the support of the individual fellowship RA 2523/1-1 from the German research organisation (DFG). F.K.~was supported by the Swedish Research Council (VR) through the Oskar Klein Centre. C.R. thanks the Albert Einstein Institute (Potsdam/Golm) for its
hospitality, particularly Lars Andersson.
F.K.~would like to thank the Institute for Gravitation and Cosmology at the University of Portsmouth for hospitality.

\appendix
\section{Derivation of the Eulerian Line Element~(\ref{ADM})}
\label{DerivationADM}

In this appendix we show how to obtain the line element~(\ref{ADM}), \ie, the Eulerian metric which results from a relativistic Lagrangian extrapolation of an approximately spherical overdensity in the deeply non-linear regime.
We use the relativistic Lagrangian perturbation theory up to second order \cite{Rampf:2013dxa} to make the non-linear frame dragging in the line element apparent. Note explicitly, that we assume a non-exact spherical overdensity; if we had assume an exact spherical symmetry, the second order terms would be exactly zero.

Before proceeding we wish to summarise the calculational steps. First, we perform a specific gauge transformation from the synchronous/comoving gauge (the Lagrangian frame) to the Eulerian gauge\footnote{In the non-perturbative treatment, the Eulerian gauge can be understood in terms of a simple ADM shift, where the shift is directly associated with the Lagrangian displacement field \cite{RampfWiegand2014}.}. The corresponding Eulerian line element will have no specified symmetry yet. By restricting the metric to an approximate spherical symmetry, we shall show that the Eulerian metric reduces to a Kerr-like metric (being however not asymptotically flat).

The synchronous/comoving line element is (summation over repeated indices is assumed)
\be
 \dd s^{2} = - \dd t^{2} + a^{2}( t )\, \gamma_{ij}(t,\fett{q}) \, \dd q^i \dd q^j \,,
\ee
with the following solution for an irrotational cold dark matter component
\begin{widetext}
\begin{align} 
\begin{split} 
 \gamma_{ij} (t,\fett{q}) &= \, \delta_{ij} \left( 1+ \frac{10}{3} \varphi \right) 
 + 3a( t )t_0^{2} \left[ \varphi_{,ij} \left( 1-\frac{10}{3} \varphi \right) 
 - 5 \varphi_{,i} \varphi_{,j} + \frac 56 \delta_{ij} \varphi_{,l} \varphi^{,l} \right] \\[2mm]
 &\quad -\left( \frac{3}{2} \right)^{2} \frac 3 7 a^{2}( t ) t_0^{4} 
 \Bigg[ 4 \varphi_{,ij} \nabq^{2} \varphi \Bigg. 
 - 2\delta_{ij} \mu_2 \Bigg] 
 + \left( \frac{3}{2} \right)^{2} \frac{19}{7} a^{2}( t )\.t_0^{4} \,\varphi_{,li} \varphi_{,j}^{,l} \,,
\end{split}
\end{align}
\end{widetext}
where we have defined $\mu_2 \defas 1/2 \big[ (\nabq^{2}\varphi)^2 -\varphi_{,lm}\varphi^{,lm} \big]$.\footnote{Here, indices are lowered and raised by the use of the Kronecker delta.}
For sake of generality we include in the appendix also the cosmological scale factor $a( t )$; for our specific case in the main text, $a$ can be viewed as a perturbation parameter (which we finally suppress in the main text). $\varphi$ denotes the cosmological potential, \ie, some initial condition (in our case with an approximately spherical symmetry) normalised at some initial time $t_0$.
The above solution can be derived by using standard cosmological perturbation theory \cite{Matarrese:1997ay}, the gradient expansion technique \cite{Rampf:2013ewa,Rampf:2012pu}, or the tetrad formalism \cite{Russ:1995eu}, and is valid for an Einstein-de Sitter (EdS) Universe.\footnote{Actually, the solutions for the Eulerian gauge hold also for a $\Lambda$CDM Universe, with only minor modifications in the time evolution coefficients \cite{RampfWiegand2014}.}
We define the Eulerian gauge by
\begin{align} \label{ADMGauge}
 \dd s^{2} &= 
- \big( 1 \big. +2 \big. A \big) \dd t^{2} 
 + 2a w_{i} \,\dd t \dd x^i + a^{2} \,G_{ij} \, \dd x^i \dd x^j\,.
\end{align}
with the spatial metric $G_{ij} = \delta_{ij}\left[ 1- 2B \right]$,
$A$ and $B$ are scalar perturbations, and, as we shall see, $\fett{w}$ contains a solenoidal and a vector part (in the non-perturbative treatment, $\fett{w}$ is just the ADM shift). We neglect tensor perturbations as they are of no importance in what follows.

The coordinate transformation is
\begin{align} 
 \label{ADMTrafo}
 &x^\mu(t,\fett{q}) = q^\mu + F^\mu (t,\fett{q}) \,,
\intertext{with} 
 x^\mu &= \begin{pmatrix} t \\ \fett{x} \end{pmatrix}, \quad q^\mu = \begin{pmatrix} t \\ \fett{q} \end{pmatrix} \,, \quad \text{and} \quad F^\mu = \begin{pmatrix} 0 \\ \fett{F} \end{pmatrix} \,,
\end{align}
where $x^\mu$ are the coordinates in the Eulerian line element~(\ref{ADMGauge}), and $q^\mu$ the one in the synchronous/comoving line element.
Explicitly, the time coordinates in both coordinate systems are identical. Thus, the above is a purely \emph{spatial} gauge transformation. The relation of the metrics is
\be
 g_{\mu \nu}(t,\fett{q}) = \frac{\partial x^{\tilde\mu}}{\partial q^\mu} \frac{\partial x^{\tilde\nu}}{\partial q^\nu} g_{\tilde\mu \tilde\nu}(\tau,\fett{x}) \,.
\ee
The metric coefficients in the Eulerian gauge in~(\ref{ADMGauge}), $\left\{ A, B, \fett{w} \right\}$, are 
unknown and we shall calculate them up to ${\cal O}(\varphi^{2})$ by using the above relation.
Truncating up to second order, the resulting relations between the space-space, space-time, and time-time part of the metrics are respectively
\begin{widetext}
\begin{subequations}
\begin{align}
 \gamma_{ij}(t,\fett{q}) &\simeq \delta_{ij} \left[ 1 -2B(t,\fett{x})\right] +2F_{(i,j)} (t,\fett{q})\left( 1-2B\right) 
\label{trunc1} + F_{l,i} F_{l,j} \,, \\[2mm]
 0 &\simeq a^{2} \left[ 1-2B \right] \frac{\partial F_i(t,\fett{q})}{\partial t}
 + a^{2} F_{l,i} \frac{\partial F_{l}}{\partial t} 
\label{trunc2} +a \,w_i(\tau,\fett{x}) + a w_l F_{l,i}
 \,,\\[2mm]
 -1 &\simeq -1 -2A(\tau,\fett{x}) +2a\, w_l \frac{\partial F_l}{\partial t}
 \label{trunc3} +a^{2} \frac{\partial F_l}{\partial t} \frac{\partial F_l}{\partial t} \,.
\end{align}
\end{subequations}
Solving Eqs.~(\ref{trunc1}--c) with an iterative technique, we obtain
for the gauge generator, \ie, the 3-displacement field up to second order
\begin{align}
 F_{i}(t,\fett{q}) &= \frac 3 2 at_0^{2} \varphi_{,i}(\fett{q}) - \left( \frac 3 2 \right)^{2} \frac 3 7 a^{2}t_0^{4} \frac{\partial_i}{{\fett{\nabla}}^{2}} \mu_2
 + 5at_0^{2} \left( \partial_i C - \partial_i \varphi^{2} +R_i \right) \,,
\end{align}
where we have defined the two terms
\begin{subequations}
\begin{align}
 C_{} &= \frac{1}{{\fett{\nabla}}^{2}{\fett{\nabla}}^{2}} \left[\frac 3 4 \varphi_{,ll} \varphi_{,mm}+ \varphi_{,l} \varphi_{,lmm} +\frac 1 4 \varphi_{,lm} \varphi_{,lm} \right] \,, \\[2mm]
 R_i &= \frac{1}{{\fett{\nabla}}^{2}{\fett{\nabla}}^{2}} \big[ \varphi_{,il}\varphi_{,mml} - \varphi_{,lli} \varphi_{,mm}
 + \varphi_{,i} \varphi_{,llmm} -\varphi_{,m} \varphi_{,mlli} \big] \,.
\end{align}
\end{subequations}
\end{widetext}
The first term denotes a purely longitudinal contribution, the latter term denotes a purely transverse contribution to the displacement and thus to the velocity field as well. Both terms are of purely relativistic origin. For the perturbations in the Eulerian line element we obtain
\vs{-2mm}
\begin{subequations}
\begin{align}
 A(t,\fett{x}) &= -\frac 1 2 at_0^{2} \varphi_{,l} \varphi^{,l} \,, \label{ADM_A1}
 \displaybreak[1]\\[2mm]
 B(t,\fett{x}) &= - \frac 5 3 \varphi(\fett{x}) + \frac 5 2 at_0^{2} \left[ \frac{1}{\nabx^{2}} \mu_2 + \frac 1 2 \varphi_{,l} \varphi^{,l} \right] \,, \label{ADM_B}
\displaybreak[1]\\[2mm]
 a\,w_i(t,\fett{x}) &= - {\cal S}_{,i}^{\rm N} + \partial_i \left[ \frac 5 3 t \varphi^{2} - \frac{10}{3} t C \right] - \frac{10}{3}t R_i \label{aw}\,,\\[-8mm]\notag
\end{align}
\vs{-1mm}
\end{subequations}
with
\begin{align}
 {\cal S}^{\rm N} &= \varphi(\fett{x}) \,t - \frac 3 2 t_0^{4/3} t^{5/3} \frac{1}{\nabx^{2}} G_2(\fett{x}) \,, \\
 \label{G2x} G_2 &= \frac 3 7 (\nabx^{2} \varphi)^{2} + \varphi_{,l} \nabx^{2} \varphi^{,l} + \frac 4 7 \varphi_{,lm} \varphi^{,lm} \,.
\end{align}
where $G_2$ is the well-known second-order EdS kernel for the velocity field at second order in Newtonian perturbation theory \cite{Bernardeau:2001qr}.
The term $A$ is the (linear) velocity of the fluid squared (generally contracted with the spatial metric which is however a third-order effect), thus denotes nothing but the Lorentz boost from special relativity. Indeed, $A$ can be written in terms of the time derivative of the displacement $\fett{F}$, as can be easily proven, $A \simeq - 1/2 a^{2} (\partial \fett{F} /\partial t)^{2}$. The scalar $B$ contains the linear initial conditions (first term in~Eq.~(\ref{ADM_B})) and some relativistic corrections. Since we are only interested here in dynamical effects, we neglect the 3-curvature in this paper and thus set the spatial 3-metric components to unity, $G_{ij} \rightarrow \delta_{ij}$ (cf.~the Minkowskian limit defined in \cite{Buchert:2012mb}).
The quantity $\fett{w}$ (see Eq.~(\ref{aw})) contains the velocity information of the fluid: $w_i = -G_{ij} a \partial {F^i} / \partial t$, where the spatial dependence on the right-hand side is \wrt~the Eulerian coordinates $(t,\fett{x})$.\footnote{As above we send $G_{ij} \rightarrow \delta_{ij}$.}
Note explicitly, that $\fett{w}$ contains not only a longitudinal velocity but also a small transverse component, cf.~the term $\fett{R}$ in Eq.~(\ref{aw}), while the metric component $A$ does only contain a longitudinal velocity component, see Eq.~(\ref{ADM_A1}). In fact, $A$ remains purely longitudinal up to third-order, so the transverse contribution in $A$ is suppressed by a factor $\varphi^{2}$.

Because of the these considerations, we write $A = -1/2 \fett{v}_{\rm L}^{2}$, $B = 5/4 \fett{v}_{\rm L}^{2}$, and $- \fett{w} = \fett{v}_{\rm L} + \fett{v}_{\rm T}$, where $\fett{v}_{\rm L}$ denotes the longitudinal part of the velocity, and the transverse velocity $\fett{v}_{\rm T}$ is suppressed by only one $\varphi$ \wrt~$\fett{v}_{\rm L}$.\footnote{The transverse velocity $\fett{v}_{\rm T}$ is proportional to $\fett{R}$.} Note that we have absorbed a scale factor in the definition in the velocities.
To derive the Eulerian metric for an approximate spherical symmetry, one then has to assume
\vs{-2.5mm}
\be
 \varphi(\fett{x}) \simeq \varphi(r) + \epsilon^{1/2}\.\varphi_{\fett{\Omega}} \,,
\ee
where the first term denotes the symmetric contribution, and the latter term a small perturbation along some given axis $\fett{\Omega}$.
The Eulerian metric is then

\begin{align}
	\dd s^{2} \simeq &- \big( 1 \big. - \fett{v}_{\rm L}^{2} \big. \big) \dd t^{2} 
 -2 \,\left( \fett{v}_{\rm L}+ \fett{v}_{\rm T} \right) \cdot \dd \fett{x} \,\dd t \nonumber\\
 &\qquad \qquad + \left[ 1
 - \frac 5 2 \fett{v}_{\rm L}^{2} \right] \dd \fett{x} \cdot \dd \fett{x}\,.
\end{align}
Note that the above relies on a Lagrangian extrapolation far into the non-linear regime, and we have nelgected the curvature metric $G_{ij} \rightarrow \delta_{ij}$. Also note that this extrapolation is formally valid for any seed with arbitrary $\varphi$.


\end{document}